\documentclass[preprint,preprintnumbers,amsmath,amssymb]{revtex4}

\usepackage{graphicx}
\usepackage{dcolumn}
\usepackage{bm}

\newcommand{\be}{\begin{equation}}  
\newcommand{\ee}{\end{equation}}  
\newcommand{\bear}{\begin{eqnarray}}  
\newcommand{\eear}{\end{eqnarray}}  
\newcommand{\ba}{\begin{array}}  
\newcommand{\ea}{\end{array}}

\newskip\humongous \humongous=0pt plus 1000pt minus 1000pt

\newif\ifdtup

\def\oldreffmt#1{\rlap{[#1]} \hbox to 2\parindent{}}

\def\figfmt#1{\rlap{Figure {#1}} \hbox to 1in{}}  
  
\def\ie{\hbox{\it i.e.}{}}	  
\def\eg{\hbox{\it e.g.}{}}	  
\def\etal{\hbox{\it et al.}}  
  
\def\Tr{\mathop{\rm Tr}}

\def\slash#1{#1\!\!\!/\!\,\,}  
\def\beq{\begin{equation}}  
\def\eeq{\end{equation}}  
\def\bea{\begin{eqnarray}}  
\def\eea{\end{eqnarray}}  
\def\half{\frac{1}{2}}  
  
\def\bq{\begin{quote}}  
\def\eq{\end{quote}}

\def\half{\frac{1}{2}}       

\def \etal {{\it et al.}\ }  
\relax  

\newdimen\tdim  
\tdim=\unitlength  
\def\bar{\overline}

\begin{document}  

\preprint{FERMILAB-CONF-05/482-T}
\vspace{1.0in}
\title{Conjecture on the Physical
Implications\\
of the Scale Anomaly}
\vskip 0.2in

\author{\vskip 0.2in 
Christopher T. Hill}

{\email{hill@fnal.gov}}

\affiliation{
 {{Fermi National Accelerator Laboratory}}\\
{{\it P.O. Box 500, Batavia, Illinois 60510, USA}}\\
\\ \\
Invited Talk delivered at the Santa Fe Institute\\
on the Occasion of the Celebration of the\\ 75th Birthday
of Murray Gell-Mann.\\ July 23, 2005\\
}%

\date{\today}
\vskip 0.2in
\begin{abstract}
Murray Gell-Mann, after co-inventing QCD,
recognized the interplay of the scale anomaly, the 
renormalization group,  and the origin of the strong
scale, $\Lambda_{QCD}$.  I tell a story, then
elaborate this concept, and for the sake of
discussion, propose a conjecture that the physical
world is scale invariant in the classical, $\hbar\rightarrow 0$, limit. 
This principle has implications for the
dimensionality of space-time, the cosmological
constant, the weak scale, and Planck scale.  
\end{abstract}

\maketitle

\section{\bf A Story}

I arrived at Caltech as a beginning graduate student in the Fall of
1972 and immediately beamed myself up to the fourth floor
of the Lauritsen Laboratory of High Energy Physics, to see what was going
on and to catch a glimpse of the great men, 
Richard Feynman and Murray Gell-Mann.
Feynman had recently attended a meeting 
in Chicago at which he had broken his knee by tripping
on a street curb. 
He was in a cast, and mostly working at home. 

On the first occasion
of a seminar in the fall quarter, however, Murray Gell-Mann showed up. 
He was dressed in a blue suit, smoking a cigar, and 
began describing the results of 
his previous sabbatical year spent at CERN.
There, working with Bill Bardeen, Harald Fritzsch and Heinrich Leutwyler, he
had put together the defining elements of what we now call QCD, the Yang-Mills 
gauge theory of the strong interactions based upon quark color
\cite{Bard1,Fri1,Fri2}. The theory had a long geneaology, emanating
from the idea of O. Greenberg of parastatistics, and the Han-Nambu model.
Gell-Mann \etal realized that the simplest assumption of a local gauge
theory of a quark color degree of freedom, commuting
with electromagnetism, made the most sense.  They didn't
call it QCD then, but rather the ``color octet vector gluon
picture.''   They had written down the lagrangian of
the $SU(3)_c$ Yang-Mills theory, 
had argued about its infrared problems and quark
confinement, had counted the number of colors, $N_c=3$,
from the anomaly governing $\pi^0$ decay, 
and understood that something had to fix the short-distance behavior
of the theory to make it consistent with the recent observations at SLAC
of electroweak scaling, as anticipated by Bjorken. 

The SLAC electroproduction data when interpreted
in terms of Bj's brilliant light-cone scaling hypothesis
\cite{bj}, represented the
first ``photographs'' of quarks moving within the nucleon at short distances.
Feynman had reinterpreted Bj's hypothesis in
the more popular language of the `parton' model \cite{feyn}. Something 
electrically neutral was seen to be carrying 
half the momentum of the proton in the infinite momentum frame,
and this was the first evidence for glue.

Throughout the fall, 1972, and winter, 1973, at Caltech 
the color octet vector gluon
picture was the main topic of discussion in Gell-Mann's 
Wednesday afternoon seminar course. 
While Gell-Mann, \etal, were coming at QCD 
from the infrared, where quarks are imprisoned within
hadrons, the forces are strong and the quarks are ``fat''
because the chiral symmetries are broken, the
SLAC data indicated that short-distance quarks were
almost massless, rattling around like little nuts in a
tin can.  During that period Murray said,
(on more than one occasion):

\begin{quote}

``Some enterprising graduate student should figure out how to reconcile
the color octet vector gluon picture with Feynman's `parton'
model.'' 

\end{quote}

This was, if properly interpreted, the most
important homework assignment of the
last third of the twentieth century. I do not know why he didn't
encode it slightly more explicitly, such as ``go calculate my
$\Psi(g)$ function'' \cite{GL} ... but such is the fog of research.
The message was clear, however: if the color octet vector
gluon picture was right, something had to happen to make quarks and gluons
act like free particles at very high recoil energy.

On the east coast there were some 
``enterprising graduate students'': David Gross,
David Politzer, and
Frank Wilczek \cite{gross,politzer}. 
I later heard a detailed account, 
over beers at the Athenaeum 
from David Politzer, as to
what led to the discovery of asymptotic
freedom in the spring of 1973.  It involved some serendipity
and the collective talents of a large number of people, as 
stated in his 2004 Nobel speech. The roles of Sydney Coleman, 
Eric Weinberg, and Feynman's famous article from 1963
on gravitation, in which Yang-Mills perturbation
theory is developed \cite{Feyn2},
were significant. 
 
Murray, upon hearing of the news of asymptotic freedom, declared
that he now considered the color octet vector gluon picture ``established.''
It was renamed
Quantum Chromodynamics (QCD), and he turned to other issues, including
a foray into gauge
unification, inventing the neutrino mass seesaw mechanism
(with Pierre Ramond and Dick Slansky \cite{GRS}), etc.  
Supersymmetry would soon arrive
on the scene, and it became
the exclusive topic of his seminar course for the remainder
of my time at Caltech (from sometime in 1975 through 1977).
Gell-Mann was an early champion of string theory.

At some point in the time frame of spring 1973 to winter 1975 an
interesting event occured.
It happened a little out of the sequence
of discussions in his course, and it was on the occasion that Murray
lectured, so I would be inclined to date it as no later than the winter
of 1975.  Present were Harald Fritzsch
and Peter Minkowski. I believe David Politzer was visiting, 
as Caltech was interested in
recruiting him, but I am a little foggy on this, and the
visitor may actually have been 
Heinrich Leutwyler.   This is the story I want to tell.

Murray that afternoon lectured about an
apparent residual ``puzzle" in understanding mass in
QCD.  Here we have ``Exhibit A'' for 
something that Coleman and E. Weinberg had 
dubbed ``dimensional transmutation''  
\cite{Coleman}.
A dimensionless number, $g_0$, which is a definite numerical value of the 
the coupling constant of QCD, is chosen
at some {\em arbitrary}, very high energy scale, $M_0$. 
Then, according the
the Gell-Mann--Low 
equation the coupling constant ``runs'' with scale $\mu$ :
\beq
\frac{d\ln{g}}{d\ln{\mu}} = \Psi(g)\;.
\eeq
Gross, Politzer and Wilczek had shown to leading
order that: 
\beq
\Psi(g) = b_0g^2\qquad \makebox{where} \qquad b_0 =
-\frac{1}{16\pi^2}\left(\frac{11}{3}N_c-\frac{2}{3}n_f \right)\; ,
\eeq 
where $n_f$ is the number of active quark flavors, e.g.,
$n_f = 6 $ including up, down, ..., top, and $N_c=3$ colors. 
Solving the renormalization group equation
we find that the running coupling constant, $g(\mu)$ perturbatively blows
up at a lower energy scale we call $\Lambda_{QCD}$:
\beq
\frac{\Lambda_{QCD}}{M_0} = \exp \left( \frac{1}{2b_0 g_0^2 } \right)\;.
\eeq
Thus, a dimensionless quantity, $g_0$, is converted into a physical scale,
$\Lambda_{QCD}$. The key point that makes this work, of course, is
the fact that $b_0$ is
negative and the theory is asymptotically free (note that
the situation is reversed
in QED, where $b_0$ is positive and the electric charge blows
up in the far UV at what we call the ``Landau pole''). 
This is the origin of mass in QCD.

It is important to realize that this has nothing directly to do with other 
mass scales in nature. This is a point that is often confused by graduate
students, who intuitively think that some other scale, like
the GUT scale or Planck scale, is somehow mysteriously feeding in
to generate the subservient strong scale. In fact, even if there were no
GUT scale or Planck scale at all, once we are
told by an experimentalist what the (nonzero) value of $g_0$ 
is at any arbitrary value of energy $M_0$, we must still have the QCD scale. 
Such is the essential miracle of
dimensional transmutation. 

Gell-Mann had been concerned with the issue of the mass
scale of the strong interactions over
the previous 20 years. He had long considered the options for
the generation of a mass scale in QCD, and had at one point
believed that it was spontaneously generated, with the concomitant
formation of a dilaton, the Nambu-Goldstone boson of a spontaneous symmetry
breaking. In other words, he was long pondering the fate
of scale invariance in QCD.

Any local field theory admits a scale current, $S^\mu$. 
Given the stress-tensor,
$T_{\mu\nu}$, which is always conserved to yield Newton's equations of
motion, $\partial_\mu T^{\mu}_\nu =0$, we can likewise construct
the moment of the stress tensor, $S_\mu = x^\nu T_{\mu\nu}$. If we then
compute the divergence of $S^\mu$ we see that:
\beq
\partial_\mu S^\mu = T_\mu^\mu
\eeq
the ``trace'' of the stress tensor.  The trace involves
the mass terms of the theory, and thus the trace is associated with
the breaking of the conservation of the scale current $S_\mu$.  
If the theory
has no mass parameters, the trace of the stress tensor will be zero
and scale symmetry will be an exact symmetry of the theory.  If the theory
has a mass scale, it represents a special scale in the theory, yields
 a nonvanishing trace, and the 
scale invariance is thus broken.

The strong interactions indeed have a mass scale
and thus must have a nonzero trace for the stress tensor.  
The trace, however, could
be a dilaton, and $\partial_\mu S^\mu \sim \Lambda_{QCD} \partial^2 \sigma $.
In this case the scale symmetry is really 
exact in the lagrangian of QCD, yet spontaneously broken by the vacuum.
So, how exactly does it work for QCD?  
How is the scale current related to the dimensional transmutation
and renormalization group origin of mass in QCD? 

At the blackboard during his lecture Gell-Mann wrote down 
the classical form of the
stress-tensor of a pure Yang-Mills theory,
\beq
T_{\mu\nu} = \Tr( G_{\mu\rho}G^{\rho}_\nu ) - \frac{1}{4}g_{\mu\nu}
\Tr (G_{\rho\sigma}G^{\rho\sigma})
\eeq
Upon tracing this we find trivially:
\beq
T_{\mu}^{\nu} = \Tr (G_{\mu\nu}G^{\mu\nu}) - \frac{4}{4}
\Tr (G_{\mu\nu}G^{\mu\nu}) = 0 \;
\eeq
Thus, classically the trace is zero and the
scale current is conserved.
Symmetries and their conserved currents are
not renormalized by perturbative quantum effects. 
Pure gluonic QCD would thus appear to be scale invariant, 
and evidently must therefore contain the dilaton. Yet, 
no experimental data supports its existence.
His lecture was inconclusive,
and it ended up in a discussion in the lecture room consisting
of Murray and a few of the
audience participants, David Politzer, Harald
Fritzsch, and
Peter Minkowski, remaining in the seminar room discussing
the puzzle. 

It so happens that the morning of this lecture I had 
been reading papers of Chanowitz and Ellis \cite{Chan1}, 
dealing with the {\em canonical conformal anomaly}.  
The authors were
mainly addressing electrodynamics, and formulating how the
conformal anomaly entered $e^+e^-$ collider physics. The main point,
relevant to Murray's discussion,  
was that $T_{\mu}^{\mu}$
generally has both a classical part, $\hbar^{(0)}\equiv 1$ representing the
defining classical input parameters, as well as an
{\em anomalous} 
part beginning at order $\hbar^{(1)}$, 
coming from quantum loops,which Chanowitz and Ellis had written as
$(R/32\pi^2)F_{\mu\nu}F^{\mu\nu}$ (where $R$ is related
to the cross-section for $e^+e^-\rightarrow \gamma\rightarrow $ hadrons,
in electron collider experiments).

I brought one of the papers into the seminar room and sat
down next to Peter Minkowski and listened.
At a lull in the discussion, I
nudged Peter, who took the paper and looked at it for
a moment.
Harald then took the paper and began to study it. Murray
at this point noticed the commotion and 
grabbed the paper away from Harald and 
began examining it. After a minute or two
Murray handed the paper back to me and said: 
``go make a copy of the first page of
this paper."  

So I went down the hall to the Xerox
machine across from Helen Tuck's office, 
and returned momentarily with the
copied first page. 
By the time I was back in the seminar 
room Murray was at the blackboard and had
written the following equation on the board
(with considerable input from Peter Minkowski,
who may have suggested the $\Psi(g)$ factor):
\beq
\partial_\mu S^\mu = \Psi(g) \Tr(G_{\mu\nu}G^{\mu\nu})\; .
\eeq
He declared that the problem was solved,
and we now understood the origin
of mass in QCD and its fundamental connection to the $\Psi$ function of
the renormalization group, and thus dimensional transmutation!  
For the next few weeks Murray seemed elated. One day he stopped by
the doorway of my office, walked in, and said with a grin, ``you realize, this
is a very interesting result!''

I thought that this new observation would be quickly written up. 
Harald and Peter continued to discuss it, but
Murray got drawn into travel and other pursuits, 
and it seemed to slip by, remaining
only as an interesting observation. It was, of course,
a reframing of something
we already knew by a different name. 
I guess I assumed that those smart guys at Princeton and Harvard already
completely understood this, and we just let it go. 
Peter Minkowski eventually wrote a nice (unpublished)
article some years later \cite{Mink}. In it he 
makes arguments similar to the original ones of 
Symanzik \cite{Symanzik}
making the context of QCD clear.
The first published study in the literature focused on QED 
and obtaining the result of eq.(7) in perturbation theory
is that of Adler, \etal, in 1977 \cite{Adler1}.
The main point in all of this is that
the scale anomaly is a general statement of the breaking of scale
invariance. The Ward identities of the anomalous current
divergence yield the renormalization group equations
when applied to the renormalized 3-point and 2-point 
functions of the theory. 
 
Murray Gell-Mann, I now believe, was perhaps the first
person to clearly see  the full connection to the scale anomaly 
that, via quantum mechanics, yields
the strong interaction scale in the real world. 
This connection is not well
known to the general community to this day. 
Even well-known and talented theoretical physicists I
have met are unaware of, and even resistant to, the fact
that the strong scale, most of the mass of the proton, 
comes mysteriously from
quantum mechanics itself. 

Therefore, befitting this occasion
I would like to expand on this notion. I would
like propose an expansive conjecture, 
even if it is merely tentative and
operational, \ie, merely for the sake of discussion -- a ``talking point.''  
This is a style
of discussion that Murray always advocated (provided it is otherwise 
sensible). 
The scale anomaly must surely have  
deeper and even more profound implications for physics than 
the remarkable origin of the strong scale. 
Let's begin, however, by summarzing how it works 
for the strong interaction.

\section{What Does it Mean?}

Anomalies are intrinsic 
effects of quantum mechanics on the symmetry structure of a field theory.

Quantum mechanics instructs us
to relate energy to frequency and momentum to wave-number
through $\hbar$. We are also instructed to compute coherent sums of amplitudes,
square them, sum over final states, and interpret 
the result as a probabilistic observable. If we set $\hbar$ to
zero, the Feynman tree-diagrams now describe the motion
of classical waves of photons and waves of electrons 
with frequencies and wave-numbers that interact
nonlinearly. In the $\hbar=0$ limit we can in principle directly measure
the wave amplitudes and we no longer
compute squares of amplitudes, and sum over final states.

Thus, at order $\hbar$ we have loop diagrams, the onset of true quantum effects.
Most of the symmetries present at the classical level of
a theory carry through into the quantum theory.  
Anomalies are exceptions. They are particular loop
effects that modify the conservation laws of the classical theory. 
These effects are fundamental and cannot be renormalized away. 

Let's consider momentarily the axial anomalies.
For example, a theory of a single Weyl spinor (\eg, a
purely left-handed relativistic spinor) coupled to
a photon does not exist because the electric current of the spinor
will have an anomaly, hence the electric
current is not conserved.
The left-handed spinor's current anomaly 
in the theory $S = \int d^4 x \;\bar{\psi}_L(i\slash{\partial}
-\gamma_\mu A^\mu -M_0)\psi_L$ is 
given in eq.(44) of Bardeen's paper, \cite{bardeen2}: 
\bea
\label{bardeen3}
\partial_\mu J^\mu & = & -\frac{1}{48\pi^2}
\epsilon_{\mu\nu\rho\sigma}F^{\mu\nu}\widetilde{F}^{\rho\sigma}
\eea
where $J_{\mu} = \bar{\psi}_L\gamma_\mu\psi_L$ 
and $\psi_L=(1-\gamma^5)\psi_L/2$. Current
conservation is essential,
since the equation of motion of electrodynamics is:
\beq
\partial_\mu F^{\mu\nu} = ej^\nu.
\eeq
where $F^{\mu\nu}$ is the antisymmetric field
strength tensor. Owing to the antisymmetry:
\beq
\partial_\nu\partial_\mu F^{\mu\nu} = 0 = e\partial_\nu j^\nu.
\eeq
The current must thus be conserved if is to be the source
term for an electromagnetic field.
Hence, a single charged Weyl fermion cannot
consistently couple to the photon through an electromagnetic current
because of the anomaly.
The anomaly destroys the symmetry of gauge
invariance, the symmetry that leads to current conservation
by Noether's theorem.  

There are many ways to solve this problem. 
Electrodynamics chooses to make the electron a ``Dirac particle.'' 
A Dirac fermion
is a pair of Weyl spinors for which the anomalies in the 
vector current cancel between the pair. The anomaly then harmlessly
resides in the ungauged axial vector current. 
The point is that electric charge conservation requires 
minimally a Dirac fermion, such as the electron in QED.

In general, if a conservation law must be enforced, as in the
case of a gauge theory which makes no sense without strictly conserved
currents, then anomalies in those current must be absent. 
We all know how this works for
the standard model, where it controls the 
quark-lepton generation structure seen in
nature. 
Axial anomalies have a fundamental topological significance. The
anomaly in even $D$ is related to the Chern-Simons term
of odd $D+1$, since $D$ can be viewed as the boundary of
$D+1$. The Chern-Simons term has
a quantized coefficient to maintain gauge invariance
in $D+1$ \cite{comment1} 
 which implies that,
apart from the coupling constant renormalization in the anomaly prefactor,
the anomaly itself is not renormalized. This is explicitly
verified in the $D$ dimensional perturbation theory,
which is the content of the 
Adler-Bardeen theorem \cite{comment2}. Axial anomalies
are thus intrinsically topological objects.

The scale anomaly is the quantum violation of the conservation
of the scale current. It is generally not topological (though
it can be linked to the axial anomaly in supersymmetric theories). 
It encodes the fundamental way in which scale invariance
is broken by quantum mechanics.
Of course, we can have fully quantum mechanical field theories that
remain scale invariant. This requires that there is a special
value of $g$, called $g^\star$ such that $\Psi(g^\star) =0$. Then
we say that $g^\star$ is a conformal fixed point. 

Conformal field theories
are of wide-ranging interest throughout physics. 
As an example, in
string theory the Weyl symmetry of the world sheet is
a $D=2$ conformal symmetry, and it
must be anomaly free since it should not matter how
one places coordinates or a metric on the world sheet.
The string target space dimensionality, \ie, the dimensionality
of spacetime for consistency with string theory, 
is selected by this criterion.
Moreover, the top quark Yukawa
coupling constant in the Standard Model is remarkably close to a 
nontrivial conformal
quasi-fixed point value \cite{cth}. 
The asymptotic freedom of
QCD implies that the theory evolves toward the 
conformal fixed point, $g^\star =0$,
as we rescale into the far ultraviolet.

We are presently interested in the generation of physical mass by
the scale anomaly. Let us display the leading factor of $\hbar$ explicitly,
and collect together the Gell-Mann-Low
equation:
\beq
\frac{d\ln{g}}{d\ln{\mu}} = \hbar \Psi(g)\;,
\eeq
and the scale anomaly equation:
\beq
\partial_\mu S^\mu = \hbar \Psi(g) \Tr(G_{\mu\nu}G^{\mu\nu})\;.
\eeq
Let us now list some observations about this system.

\vskip 0.2in
\noindent
{\em (i) The Gell-Mann--Low Renormalization Group is
Equivalent to the Scale Anomaly}
\vskip 0.2in
 
The Gell-Mann--Low renormalization group is one of 
the earliest instances of recursion in
the scientific literature \cite{GL}. 
The equation tells us how
the theory continuously morphs from one scale to another, but 
in a way that
is independent of the scale itself, depending only upon the structure
of the theory at any given scale, 
which dictates the form of $\Psi(g)$, and its coupling constant $g$. 
Thus, as we move up or down in energy, the theory makes a copy of
itself with a new value of $g$ in a self-similar fashion.

The scale anomaly and the Gell-Mann--Low equation
are essentially equivalent. The scale anomaly
may be viewed as slightly more general than the RG equation
since the renormalized Ward Identities of the 3-point function
can be massaged into the form of the RG equation.
Nonetheless we can reverse the procedure
and even ``derive'' the scale anomaly from the Gell-Mann--Low
equation on the back of an envelope if we hand-wave a bit.
Consider a scale transformation of the lagrangian:
\beq
L = -\frac{1}{2g^2}\Tr G_{\mu\nu}G^{\mu\nu}\;.
\eeq
Treat $g$ as a running coupling constant
and let $Q=\int d^3x S_0$ be the scale charge. If we commute
$Q$ with an operator $Y(x)$ we generate an infinitesimal scale transformation 
$dY/d\ln\lambda$, where 
$Y\rightarrow \lambda^d Y(\lambda x)$. Thus, given that
the classical trace of the stress tensor is zero, we have 
$[Q,\Tr G_{\mu\nu}G^{\mu\nu}]=0 $. This can be preserved at loop
level provided we sweep all scale dependent
renormalizations into $g^2$.
However, we then have that:
\bea
[Q, L] & = & \frac{\partial}{\partial\ln\lambda}L 
\nonumber \\
& = & 
-\frac{\partial}{\partial\ln\lambda}\left(\frac{1}{2g^2}\right)
\Tr G_{\mu\nu}G^{\mu\nu}
 =  -2\hbar\Psi(g)L\;.
\eea
Passing to canonical
normalization the last term is $\hbar\Psi(g)\Tr G_{\mu\nu}G^{\mu\nu}$
which is the usual scale anomaly.

\vskip 0.2in
\noindent
{\em (ii) The Scale Anomaly Naturally Generates Hierarchies}
\vskip 0.2in

Let's reemphasize the fact that the mass scale generated in QCD comes
from quantum mechanics. It does not ``trickle down'' from some
higher energy scale, such as $M_{Planck}$. However, we can 
imagine that there is physics at the Planck scale, \eg, 
string theory, that actually sets the value of the strong coupling constant.
For example, perhaps some enterprising graduate student will one day
show that $\alpha_s(M_{Planck}) \approx 1/4\pi^2 + ...$.  Then 
the renormalizaton group 
automatically runs the coupling into the infrared and establishes a large
hierarchy of $\Lambda_{QCD}/M_{Planck}$.  If the theory is successful, the
correct value of $\Lambda_{QCD}/M_{Planck} \approx 10^{-20}$ is predicted by:
\beq
\frac{\Lambda_{QCD}}{M_{Planck}} = \exp \left( -\frac{1}{8\pi \hbar b_0 
\alpha_s(M_P) } \right) \;+\;\makebox{(higher order corrections)}
\eeq 
It is a stunning aspect of the renormalization group
that a twenty order of magnitude hierarchy can naturally
be generated by a normal perturbative input parameter, 
$\alpha_s(M_{Planck}) \sim 10^{-1}$.
In fact, we know of no other way to do it.
The detailed explanation of the origin of the hierarchy between the weak scale,
$G_F^{-1/2}$, and the Planck scale, remains a mystery. It is tempting
to believe that a similar mechanism may underlie the electroweak hierarchy.

\vskip 0.2in
\noindent
{\em (iii) The Custodial Symmetry of the Hierarchy is
Classical Scale Invariance!}
\vskip 0.2in

In the 1970's 't Hooft proposed a general rule concerning hierarchies
\cite{hooft}. Namely,
when we have a hierarchy of two physical quantites, such as $a/b << 1$,
then in the limit that $a/b = 0$ there will always be 
a ``custodial symmetry'' that
maintains the special value  $a/b = 0$ to all orders of perturbation theory.

For example, consider the $e$-$\tau$ mass hierarchy in the standard model,
$m_e/m_\tau \sim 10^{-4}$. In the limit that $m_e/m_\tau = 0$ we have the 
$U(1)_L\times U(1)_R$  chiral
symmetry (modulo a harmless anomaly) of the electron. 
This chiral symmetry is then maintained to all orders 
of perturbation theory, and the nonzero value
of $m_e$ is not regenerated.

We cannot
make $b_0\propto -(11N_c - 2n_f) \rightarrow 0$ 
by tuning integers $n_f$ and $N_c$. The issue of taking
the limit must actually apply in
pure QCD with $n_f=0$, and $N_c\rightarrow 0$ is then meaningless. 
Even if
we could engineer $b_0=0$ at order $\hbar$ we would still have running at
order $\hbar^2$, {\em etc}.   
The ``Banks-Zaks fixed
point'' attempts to cancel the $O(\hbar)$ against the $O(\hbar^2)$
term, but this largely is a model building
tool, employed \eg, in walking technicolor (see \cite{simmons}), 
but is not a useful mathematical lever for taking the limit. 
We could take $\alpha_s(M_P)
\rightarrow 0$, the conformal fixed point, but then we lose 
the interactions of QCD at all scales. 

If we would like to have a limit
in which the structure and interactions of the theory are intact
the only parameter at our disposal to vary remains $\hbar$.  We see
that in the {\em classical limit} $\hbar\rightarrow 0$ the hierarchy
$\Lambda_{QCD}/M_{Planck} \rightarrow 0$ (assuming $M_{Planck}$ is held fixed!)

Thus, we can view the custodial symmetry of the strong hierarchy as 
a classical limit in which all anomalies are turned off and the
scale symmetry is exact.

\section{A Conjecture}

The notion of a classical symmetry
as the custodial symmetry of a quantum mechanical hierarchy seems to be
arguably at work in QCD. We now wish to amplify this notion, and we begin by
suggesting a talking point conjecture:

\begin{quote}

All mass scales in physics are intrinsically quantum mechanical and
associated with scale anomalies. The $\hbar\rightarrow 0$ limit of nature is
exactly scale invariant.

\end{quote}

In fact, since the notion of classical scale invariance 
as the custodial symmetry
of the QCD mass scale makes sense,
then it may seem absurd that other mass scales be intrinsically classical. 

On the other hand, this limit may make no sense for many theories,
including some of our favorites. String or M-theory 
has a classical input parameter and we would be challenged to
find a quantum mechanical origin for this. However, string theory
is intrinsically quantum mechanical
and the limit $\hbar\rightarrow 0$ may be meaningless.
In particular, theories that are defined by duality in a fundamental
way may not have a meaningful classical limit. In the modern
parlance, ``duality'' refers to physical states 
that have properties (masses, charges, etc.)
that scale as $1/\hbar$, and are usually topological,
and often in one-to-one correspondence
with states whose properties may scale as $1$ or $\hbar$. 
Thus, magnetic monopoles, whose
magnetic charges scale as $1/e\hbar$ are not compatible with the 
$\hbar\rightarrow 0$ limit. However, we'll argue below that, since
these objects are intrinsically topological and are 
associated with the mass scales
of spontaneously broken symmetries, objects like the 't Hooft-Polyakov 
magnetic monopoles 
unwind into nothingness as these symmetries are restored
in the scale invariant limit. 

It isn't easy to contemplate all of the things that happen if
we try to take $\hbar\rightarrow 0$ in reality. What we really mean, 
in a weaker sense, by this conjecture is the statement that:
\beq
T_\mu^\mu ={\cal{O}}(\hbar)
\eeq
There are no classical or ${\cal{O}}(1)$ parameters in the 
trace of the stress tensor in a 
perturbative power series expansion in $\hbar$. String theory,
in a larger sense, intrinsicaly
involving quantum mechanics may be viewed as
consistent with this hypothesis.

Let us now briefly summarize the naive consequences of this conjecture. 
The arguments will all be hand-waving, and 
there would be much more to do to develop this hypothesis 
beyond what I will present in this brief discussion. 

\vskip 0.2in
\noindent
{\em (i) A selection rule explaining why we live in $D=4$?}
\vskip 0.2in

Consider any one of the Yang-Mills elements of the
standard model, \eg, QCD, in $D$ dimensions. The 
classical stress tensor has the
same form, but taking the trace yields:
\beq
T_{\mu}^{\nu} = \Tr G_{\mu\nu}G^{\mu\nu} - \frac{D}{4}
\Tr G_{\mu\nu}G^{\mu\nu} \;.
\eeq
This must be zero by hypothesis, hence $D=4$.

It is well-known that the
couplings constants in $D\neq 4$ carry mass dimension. 
The Gell-Mann--Low
equation is modified, having the classical part:
\beq
\frac{d\ln{g}}{d\ln{\mu}} = \half (D-4)+\hbar \Psi(g)
\eeq
The first term on the {\em rhs} leads to power-law
running of the coupling constant.
Our conjecture, however, provides a simple selection rule for the
dimensionality of space-time. The classical part of
the above equation must vanish. And, it agrees with experiment.

Does this imply that extra compact dimensions cannot
exist?  They can, but they would have to be
quantum mechanical in origin and would have an effective radius 
associated with the scale provided by a scale anomaly.
We do know a simple way to construct fake effective extra dimensions
in this manner. 
This can be done by way of ``deconstruction,'' \cite{decon}, which is a lattice
of QCD-like theories wired together in a  sequence so that an effective
tower of KK-modes appears in the spectrum. It is really none other than
a Wilson style lattice description of an extra dimension, where the Wilson
links are chiral fields coming from QCD-like condensates. 
Such a system could be consistent with our
conjecture, being classically scale invariant. 
In the $\hbar\rightarrow 0$ limit the theory would
fall apart into a large number of classical massless Yang-Mills
theories. 
In fact, since there is no geometrical principle at work here, 
the renormalization group itself must serve to define the lattice 
of an effective quantum extra dimension. This has solutions that are
geometrical,
but also fractal ones as well \cite{cth1}, 
where the number distributions of KK-modes 
with energy can have a nonintegral exponent:
\beq
N(E) = (E/M)^{\hbar({D-4})}
\eeq 
This bears some resemblance to QCD with $(D-4)/2 \sim \Psi(g)$
and $\Lambda_{QCD} \sim M$, since it is the pattern structure
of a renormalization group solution. [The ``renormalization group as a
substitute for geometry'' 
seems to me to be yet another
hypothesis worthy of further consideration elsewhere (see \cite{cth1})].

\vskip 0.2in
\noindent
{\em (ii) Infrared cancellations, topology and
tunneling.}
\vskip 0.2in

It is required that the
probabilisitic quantum measurement theory
no longer hold in the $\hbar=0$ limit. 
Dispersion relations relate probabilistic observables
to (imaginary parts of) loop diagrams, hence as
$\hbar\rightarrow 0$ so too go the probabilistic observables.
Moreover, familiar 
infrared cancellations in QED between collinear
emission tree amplitudes, which are $O(\hbar^0)$ in amplitude,
are known to produce $O(\hbar^1)$ singularities in 
the integral over phase space.
In the full quantum theory these cancel against infrared $O(\hbar^1)$ loops
\cite{Yennie}. Thus, we may view the $\hbar\rightarrow 0$ limit, which
eliminates the loops, as also
forcing us to abandon the probabalistic
interpretation of quantum mechanics with the sum over final states. 
This isn't surprising since the classical theory
selects a unique outgoing state for any particular incoming
state. Any singularities are now artifacts of the quantum
calculational method and no longer part of the physics. The sum of
tree diagrams yield the perturbation series of the Fredholm theory
of the nonlinear wave theory for any particular incoming classical
state.  

Note that WKB tunneling is also suppressed, faster than any power,
as $\hbar\rightarrow 0$. Thus instantons (tunneling) 
never occur in the classical
theory. As mentioned above,
states involving duality, typically topological objects like
monopoles, skyrmions, disappear in the $\hbar \rightarrow 0$ limit
as a consequence of scale invariance.
These typically involve the topological
current, $\epsilon_{\mu\nu\rho\sigma}
\Tr (U^\dagger\partial^\nu U  
U^\dagger\partial^\rho U U^\dagger\partial^\sigma U) $ where $U^\dagger U =1$.
Without the latter unitary constraint there is no conserved topological
current, and
in most cases, such as monopoles and skyrmions, this arises from a nonlinear
sigma model in which a field satisfies $\Phi^\dagger \Phi = v^2$.
Thus in the $\hbar\rightarrow 0$ limit,  $v\rightarrow 0$, and therefore
a scale invariant world won't contain as
much topology. The dual toological states furthermore
help to lock scale invariance to the $\hbar \rightarrow 0$ limit,
enforcing our conjecture. 

Of course, these are naive considerations.  As we've
noted, in theories in which
a symmetry, such as the inversion symmetry of $M$-theory, is
fundamental and maps dual states into anti-duals, etc., the
$\hbar\rightarrow 0$ limit may be meaningless.

\vskip 0.2in
\noindent
{\em (iii) The cosmological constant is classically zero.}
\vskip 0.2in

The cosmological constant is a term
in the stress tensor of the form $T_{\mu\nu} = \Lambda g_{\mu\nu}$.
Our  conjecture requires $T_\mu^\mu = 0$, hence $\Lambda=0$.
This does not, of course, solve the problem of the cosmological constant, but
it eliminates the usual classical aspect of it. The situation is akin to
supersymmetry, in which exact SUSY also implies vanishing vacuum energy. 

One thing is fairly clear: whatever zeroes $\Lambda$, leaving a possible
ultra-small residual component with $\Omega \approx 0.6$, 
must be mechanism that
can be understood within the context of any given
scale anomaly. Thus, for example, we must be able to understand
the zeroing mechanism within the context of QCD alone. 

Indeed, the scale anomaly of QCD will produce a cosmological
constant as:
\beq
\Lambda \sim \frac{1}{4}<0|\Psi(g) \Tr(G_{\mu\nu}G^{\mu\nu})|0> \sim
\Lambda_{QCD}^4.
\eeq
One potential remedy for this would be the presence of an
dilaton, and we return to this below. 

\vskip 0.2in
\noindent
{\em (iv.a) The QCD scale is generated by quantum mechanics.}
\vskip 0.2in
This is the basis of the conjecture.

\vskip 0.2in
\noindent
{\em (iv.b) The conjecture may be testable in the weak interactions}
\vskip 0.2in

The original Weinberg-Susskind idea of Technicolor was largely
motivated by the naturalness of the QCD hierarchy. It would produce
the weak scale by way of a scale anomaly in the manner of QCD.
Something like Technicolor would confirm the conjecture.

The first forays into this, \ie, the various
incarnations
of Technicolor in which fermion masses are treated as small,
have largely failed. 
Technicolor models that do not reckon with the heaviness of the top quark
(and even the charm and bottom quarks) are largely ruled out. 
Conversely, models that treat the top quark as part of the dynamics
have enjoyed some success, and are still viable \cite{simmons}. 
It is certainly {\em
not true} that these models are ruled out, and they are
in principle completely compatible with SUSY.  Indeed, most
recent efforts to ameliorate the fine tuning problems inherent in
the MSSM are pushing that theory in a less perturbative direction with
more emphasis on the third generation. Moreover, SUSY breaking is generally
approached as  a dynamical mechanism, consistent with
the trace anomaly conjecture.

There are four classes of viable dynamical weak scale models 
and some are 
testable soon (possibly at Tevatron, surely at LHC).  One is to
combine Technicolor with a dynamics (Topcolor) that can accomodate
the third generation heavy masses. This is built out of precursory models
of Eichten and Lane in which multi-scales of condensates are involved. 
These models are thus complex \cite{Lane}. These
models offer lots of targets to experiment. 
Perhaps most intriguing would be something like a b-tagged dijet mass
excess in the $W+2j$ channel at the Tevatron. Since Run I we have
seen a modest excess there, but nothing is yet conclusive. 

My personal favorite model is the Top Seesaw \cite{cth2}. 
This is motivated by
the fact that if the top quark weighed in at 600 or 700 GeV, 
then there would be
absolutely no doubt that we have a natural strong dynamics at the weak scale,
and we would expect dimensional transmutation generated it. In fact,
we can easily engineer such a model using Topcolor.  Chivukula,
Dobrescu, Georgi and I considered implementing the
Gell-Mann--Ramond--Slansky seesaw \cite{GRS}, which adds new
ingredients that can rotate the physical mass down to
its 175 GeV observed value. 
This then lifts other objects (the $\chi$ quarks) up
to the few TeV scale. The additional ingredients must all have dynamically
generated scales, but this appears possible and yields various heavy PNGB's.

The Top Seesaw was preprinted in 1998 and was DoA (Dead on Arrival) as the
LEP S-T error ellipse at that time was disconcordant with the theory 
by about 5-$\sigma$. To our delight, in 1999 LEP recalibrated its beam energy, 
and the error ellipse
lurched to the ``northwest'' on the plot. The Top Seesaw was thus ruled in
at the 2-$\sigma$ level.
I like to say that we predicted this, but pundits say that we
fine--tuned our theory. This strikes me as slightly acausal. 

Little Higgs theories provide another interesting approach that
attemptsto treat the Higgs boson as a Nambu-Goldstone boson. Here
the strong dynamics occurs at $\sim 10$ or $20$ TeV, making a chiral
condensate, leading
to ``mesons'' (Nambu-Goldstone bosons) 
that are typically weak singlets (like $\eta$), 
doublets (like $K$)
and vectors (like $\pi$) \cite{little}. Then at the $\sim 1$ TeV scale there is
top dynamics, similar to the Top Seesaw in structure
that generates a Coleman-Weinberg potential in which $K$
develops a VEV and becomes the Higgs boson. These models are thus semi-natural,
postponing the new strong dynamics upward in energy by an order of magnitude. 
They are challenging to engineer because one must forbid an
induced $\pi^0$ VeV (T-parameter constraint)
and still have the global minimum of the Coleman-Weinberg potential 
at $v_{weak} << 1$ TeV. Evidently only larger chiral symmetries with
certain tensor representations suffice (what Georgi refers to as
``kissing mexican hats''). 

An intriguing rennaissance of dynamical models using
the neutrino sector, largely due to Appelquist and Shrock, 
is also quite appealing \cite{Appelquist}. The role of neutrino
mass in dynamical models is an interesting contemporary issue.
Thus the weak scale may involve some interesting new dynamical
physics.

\vskip 0.2in
\noindent
{\em (v) Scale invariant
gravity in the $\hbar\rightarrow 0$ limit?}
\vskip 0.2in

Here I will briefly mention an older, non-stringy approach
to gravity, which is largely motivated by analogy to QCD.
This attempts to directly derive the Planck scale in a manner
similar to the strong QCD scale.  Whether this is string theory
in some kind of disguise is unclear, if not dubious. It in fact it suggests
that the continuum of space-time procedes to much shorter distances
than the Planck scale.
There is, however,  a dramatic phase transition in physics within
these theories at the Planck scale.

Indeed, there is an extensive literature of ``quadratic gravity,'' 
begining with the conformally-invariant
gravity of Weyl \cite{Weyl}, and subsequent incarnations
due to Adler \cite{Adler2}, Mansouri \cite{Mansouri}, Tomboulis \cite{Tom}, 
\etal, as well as 
detailed analyses of renormalizability 
by Stelle \cite{Stelle}, {\em et.al.}
These seem optimally related to our present conjecture,
though much of the interest in this venue halted with the rise
of string theory in the early 1980's. 

The key question is how the $\sqrt{-g} M_{Planck}^2 R$ term 
could be generated
dynamically by  a quantum scale anomaly?  Classical scale
invariance is essentially the 
logic underlying some of the aforementioned
work, \eg, in particular Adler's formulation \cite{Adler2}.
He  implements the classical scale
invariance by choosing dimensional regularization
as a definition of quantum loops. 
One can sketch a scenario based upon
these papers, in particular the nice work of Tomboulis \cite{Tom}. 

The method here is to imitate QCD.
Since in QCD, at high energies or in the
$\hbar\rightarrow 0$ limit, we have scale invariance, therefore
by analogy we seek a starting point for pure gravity
that is scale invariant at high energies.
Such a theory can be built of scale
invariant terms that define ``quadratic gravity,''
\beq
\label{one}
\frac{1}{h_1^2}\sqrt{-g}R^2 + \frac{1}{h_2^2}\sqrt{-g}R_{\mu\nu}R^{\mu\nu}
+ \frac{1}{h_3^2}\sqrt{-g}R_{\mu\nu\rho\sigma}R^{\mu\nu\rho\sigma}
\eeq 
The $\sqrt{-g}R_{\mu\nu\rho\sigma}R^{\mu\nu\rho\sigma} $
can be written in terms of the first two operators if we set the
Gauss-Bonnet invariant to zero. The Gauss-Bonnet term is a topological
index that takes on discrete values for metrics that are not
continuously deformable to flat-space, (much
like the Pontryagin index, $\Tr G\tilde{G}$ of
Yang-Mills). For simplicity we do not include it.
Moreover, if one demands an identically zero trace
for the classical gravitational stress-tensor, (derived by differentiating
eq.(\ref{one}) wrt $g_{\mu\nu}$), then one is led to a unique
lagrangian which involves the Weyl tensor,
 $\sqrt{-g}C_{\alpha\beta\gamma\delta}C^{\alpha\beta\gamma\delta}$.
This can be written, following Tomboulis \cite{Tom}, as:
\beq
 \label{two}
\frac{1}{h^2}\sqrt{-g}(R_{\mu\nu}R^{\mu\nu}-\frac{1}{3}R^2 )
\eeq
This is taken to be our high energy theory, at scales well above
the Planck mass. There is no
$\sqrt{-g}M_P^2 R $ term yet, as such will be generated in analogy
to the QCD mass scale through a scale anomaly.

Now, there are immediate apparent problems at the outset with this theory.
If we view the metric as the fundamental degree of freedom,
this Lagrangian has quartic derivatives, and
the  graviton propagator has the form,
$\sim 1/p^{4}$. Such a quartic propagator can be
viewed as the limit:
\beq
\lim_{m^2\rightarrow 0} \frac{1}{m^2}\left( \frac{1}{p^2} - \frac{1}{p^2 +m^2}\right)
\eeq
The second term on the {\em rhs} has a negative residue. Thus the
theory can be considered as the limit of a massless graviton and a massive
ghost field as the ghost mass becomes small.
The addition of the $M_P^2\sqrt{-g}R$ term would modify the
propagator as $\sim 1/(p^2 -p^4/M_P^2)$, thus lifting the ghost
mass, but not changing the negative norm of the ghost states.
Ghosts, of course, spoil unitarity if they are asymptotic in- or out-
states in the $S$-matrix. Thus, our high energy gravity theory
as a quantum theory is somewhat sick.

This may be a harbinger of other nonperturbative sources
of unitarity violation in gravity. For example, the creation
by quantum fluctuations in the vacuum, \eg, the formation of mini-black-holes
is also a putative unitarity problem for gravity. 
Tomboulis, exploiting old ideas of Lee and Wick, argues
that this problem can be ameliorated if the massive ghost is sufficiently
unstable that it doesn't really enter asymptotic in or out states \cite{Tom}.
However, the problem is then swapped for fluctuations in causality
at the Planck scale.  

Irrespective of ghosts, it has been shown by
several authors that this theory is renormalizable \cite{Stelle}!
The $1/p^4$ sufficiently softens the UV structure of Feynman amplitudes
and the scale symmetry, broken only by the trace anomaly, enforces
a multiplicative renormalization of the action. 

Tomboulis has proposed an intriguing idea that simply
imitates the renormalization group behavior of the theory
in a manner similar to QCD. The idea is to incorporate matter, in
particular a large number $N$ of Weyl fermions, added to
the lagrangian,
\beq
\sqrt{-g}\sum_{i}^{N}\bar{\psi}_i\slash{D}\psi_i
\eeq
where $\slash{D}$ is a gravitational (or otherwise) covariant derivative,
which is implemented in the vierbein formalism. 
In the large $N$ (fermion bubble approximation)
limit we can obtain the Gell-Mann--Low equation
for the gravitational parameter, $h$. From Tomboulis we have \cite{Tom}:
\beq
\frac{d\ln h}{d\ln\mu} = b_0h^2 \qquad b_0 =-\frac{\hbar N}{160\pi^2}
\eeq
Remarkably, fermion loops cause the theory to be
asymptotically free, an essential completion of the analogy
with QCD.  The RG running implies that there will thus be a scale anomaly,
which we can likewise infer from Tomboulis' work:
\beq
\partial_\mu S^\mu = \frac{\hbar Nh^2}{160\pi^2}
(R_{\mu\nu}R^{\mu\nu}-\frac{1}{3}R^2 )
\eeq
(here we have passed to canonical normalization of the graviton; I do
not know what the $\psi$ function for the pure gravity part of the theory
is; this theory may always be asymptotically free due
to the attractive nature of gravitation).
 
This picture is quite QCD-like. It implies that there will be a scale,
generated by dimensional transmutation, at which $h$ blows up, given
a measurement of $h=h_0$ at some arbitrary higher energy scale $M_0$:
\beq
\frac{\Lambda_{Planck}}{M_0} = \exp(-160\pi^2/Nh_0^2)
\eeq
The dynamics of the theory is thereby modified
at the scale $\Lambda_{Planck}$. Though a leap
of faith, one naturally expects that normal gravity is generated
at this scale. I say leap of faith, because the matching may be complicated,
and one would
expect some analogue vestige of quark-gluon confinement to occur -- the low
energy graviton may be a boundstate of confined high-energy degrees
of freedom. 
Adler \cite{Adler2} gives explicit formulae for
the induced cosmological constant and 
dynamically generated Planck mass. Naturally,
we expect $M_{planck}\sim \Lambda_{Planck}$ and $\Lambda_c\sim \Lambda_{Planck}$
in this scheme.

Perhaps one should  not be  {\em a priori} bothered by
the unitarity issues, as they may in fact be 
subtle effects of quantum gravitation that have not
been seen by low energy experiment. 
The issue is in part cosmological. 
It is not hard to see that Weyl gravity has a scale invariant
Friedman-Robertson-Walker solution in a scale invariant radiation dominated
universe (as would be the case for $N$ Tomboulis fermions). The
Hubble constant scales as $H= \dot{a}/a \sim 1/t $. This is presumably
the solution at high energies, above the phase transition into
normal Einstein gravity at which $M_{Planck}$ forms. Thus, it
would be of interest to revisit the unitarity issue, \ie, the pole
structure about this or other background geometries. 

Perhaps another way to circumvent the problem of unitarity violation
is to view this as a theory in which the Christoffel symbol, rather
than the metric, is the
dynamical degree of freedom.  Mansouri proposed \cite{Mansouri} that 
the Christoffel symbol can
be viewed as the Yang-Mills gauge field of a local GL(4,R) symmetry. 
The Christoffel symbol is related to the metric
in the usual way. The relationship is not a dual relationship, but
does bear slight resemblance to the relationship of the 
axion to the Kalb--Ramond
field. One path might be to intepret $\sqrt{-g(x)}$ as the
trace of a Wilson line that parallel transports a boundary 
metric $g_{\mu\nu}(0)$
from $0$ to $x$ using the
Christoffel symbol, 
$g_{\mu\nu}(x)\sim \exp(\int^x_0 \Gamma^\mu_{\mu\rho}dx^\rho)$. 
This allows one to write an action that depends only upon 
$\Gamma^\mu_{\nu\rho}$, but has become a nonlocal theory 
of Wilson lines coupled to
local fields. Perhaps this 
somehow revisits string theory and holography.
It would
be interesting if we could
interpret the high energy theory as purely affine, and that there
is no fundamental metric. Then gravity may be ``emergent.''

I have selected this
scenario to illustrate a gravitational mimic of QCD, 
and have foregone  addressing the very large number of residual issues.
I have no idea if this picture can be reconciled with string theory
in some fundamental way to argue that the classical string
scale is also emergent. It seems that it has some interesting
things to say, and is worthy of further elaboration. various
questions arise, \eg, can inflation occur at scales above $M_{Planck}$?

\vskip 0.2in
\noindent
{\em (vi) The dilaton?}
\vskip 0.2in

We see the characteristic problem of a cosmological constant
emerging in any strong dynamics when $\hbar\neq 0$ \cite{Adler3}.   
It seems to me that the only
resolution to this must involve
the existence of an additional degree of freedom,
a dilaton, $\sigma$, which participates in the scale anomaly
itself:
\beq
\label{three}
\partial_\mu S^\mu = (\makebox{all scale anomalies}) - f_D\ D^2\sigma -V'(\sigma)
\eeq
This is analogous to the axion in the axial anomaly:
\beq
\label{four}
\partial_\mu A^\mu = (\makebox{all axial anomalies}) - f_a\partial^2 a - V'(a)
\eeq
Such a dilaton implies that the world is truly
scale invariant when $\hbar \neq 0$. It in some sense
implies that $\hbar\neq 0$ is a spontaneous breaking of 
a classical world.

Such a dilaton 
can be introduced ``evanescently'' by way of
dimensional regularization by incorporation of
factors in loops like $\exp(\epsilon \sigma/f_D)$. This maintains
quantum scale invariance as one goes to $\epsilon = D-4$ dimensions.
This will generate an equation such as eq.(\ref{three})
with an induced kinetic term for the dilaton as well.
The ``evanescent'' dilaton is purely quantum mechanical,
associated only with loops and its couplings absent in the classical
theory. It may represent local small changes
in space-time dimensionality, and even lead to a 
dynamical interpretation of $\hbar$ itself.

The point is that gravity is supposed to see the entire {\em rhs} or
eq.(\ref{three}) (while, \eg, the proton only sees the QCD scale anomaly from which
it gets its fixed mass).
The {\em rhs} of eq.(\ref{three}) is now zero by the dilaton equation of motion.
By analogy to the case of axions, eq.(\ref{four})
governs the putative physics of an axion detector. The detector
produces the $F\tilde{F}$ source term, and  radiates
energy away into axions. 
In the case of the dilaton, we would hope that
this provides a relaxation mechanism for the
cosmological constant, perhaps even a small relic one if the dilaton
is slightly off mass-shell in the universe today. 

The idea of using a massless dilaton to eliminate $\Lambda_{cosmological}$ 
can only make sense in a world in which we can take $f_D >> M_{Planck}$.
Otherwise, in the case that $f_D/M_{Planck} \leq 10^{-3}$
the dilaton produces long-range scalar corrections to gravity
that can be ruled out by experiment limits.  The existence of scales
beyond $M_{Planck}$ removes this hurdle to incorporating a dilaton.

\vskip 0.2in
\noindent
{\em (vii) Softly broken classical scale invariance}
\vskip 0.2in

Much of this discussion was motivated
by a comment a decade ago
of Bill Bardeen who pointed out that, in the Standard Model, 
the Higgs boson mass 
can be viewed as a ``soft classical scale symmetry'' breaking parameter
\cite{Bardeen}.
This means that, if the Higgs mass were the only physical scale
near to and above the weak scale, then it is {\em technically natural}.
This is not the case for the usual Standard Model, which has
the $U(1)$ Landau pole, but one could imagine a modifed
Standard Model in which it holds modulo gravity.

Bardeen had in mind scale breaking only by way of 
logarithmic effects in loop
integrals, which lead to the scale anomaly. 
This was refutation of something we hear often in conference talks about SUSY.
Namely, it is often stated that ``the Standard Model Higgs mass is subject 
to additive quadratic
divergences in field theory, and is therefore destabilized and unnatural.''
Bardeen points out that technically 
the only source of quantum scale breaking is the scale
anomaly, and this is a $d=4$ operator, 
and cannot include the $d=2$ Higgs mass term. The Higgs mass term
is a classical input to the standard model, and thus can be
viewed as a soft classical scale breaking term,
provided we somehow switch off any Landau poles.

Thus, any additive quadratic divergences are 
an artifact of the choice of a
momentum space cut-off, or a Pauli-Villars regulation procedure. 
One sees this happen in any regulator
that violates classical scale invariance. However, if one truly implemented
classical scale invariance as a limiting symmetry of the theory, with the
Higgs mass as a soft breaking parameter, such effects would
be removed by the Ward identities of the theory.  

Put another way, if we were to consider massive scalar electrodynamics,
and we compute with a regulator that respects classical scale invariance
(\eg, dimensional regularization, which may be the only one) then we
find that the scalar mass term is multiplicatively renormalized.
Essentially, classical scale invariance can provide, 
in principle, the same degree
of technical naturalness that SUSY does for the weak scale.

\section{Conclusions}

We have argued that, for the entire physical world,
$T_\mu^\mu = O(\hbar)$, and has no surviving ``classical'' 
term as $\hbar\rightarrow 0$.
As a principle, this would be a powerful constraint on the world. 
There is an appealing naturalness
to the idea of all mass arising from scale anomalies, a purely quantum
mechanical origin, that is satisfying. 
QCD may be sufficient evidence
of this principle. 
String theory may, through its intrinsic duality,
provide an exceptional and  nontrivial realization 
of the principle. 

Murray Gell-Mann, to my reckoning, was the first person to close
the loop and appreciate how the
fundmental scale of the strong interactions is intimately tied
to the the scale anomaly, the renormalization group and quantum mechanics.
He was also a great teacher and mentor.  He made Caltech's HEP theory group
a stimulating place in which to work and study. 
We are here, in the spectacular surrounds
of the Santa Fe Institute, for the celebration 
of his 75th birthday, and to salute him.  
To this, may the lights of Babylon burn brightly 
for many years to come! 

\vspace{0.5 in} 
I thank Steve Adler, Bill Bardeen and Peter Minkowski
for useful discussions.

\end{document}